\documentclass[%
 reprint,
 amsmath,amssymb,
 aps,
]{revtex4-1}

\usepackage{graphicx}
\usepackage{dcolumn}
\usepackage{bm}

\begin{document}


\title{On shock waves from the inhomogeneous Boltzmann equation}
\thanks{\bf Our paper is dedicated to the memory of Carlo Cercignani.}%

\author{Yves Pomeau}
 \email{yves.pomeau@gmail.com}
 \affiliation{LadHyX - Laboratoire d'hydrodynamique, Ladhyx, Ecole Polytechnique, Palaiseau, France.}
\author{Minh-Binh Tran}%
 \email{minhbinht@mail.smu.edu}
\affiliation{Department of Mathematics, Southern Methodist University, Dallas, TX 75275, USA
}%


%

\date{\today}

\begin{abstract}
We revisit the problem on the inner structure of shock waves in simple gases modelized by the Boltzmann kinetic equation. In  \cite{pomeau1987shock}, a self-similarity approach was proposed  for infinite total cross section resulting from a power law interaction, but this self-similar form does not have finite energy. Motivated by the work of Pomeau, Bobylev and Cercignani started the research on the
rigorous study of the solutions of the spatial homogeneous Boltzmann equation,  focusing on those which do not have finite energy 
 \cite{bobylev2002self,bobylev2003eternal}. However, infinite energy solutions do not have physical meaning in the present framework of kinetic theory of gases with collisions conserving the total kinetic energy. In the present work, we provide a correction to the self-similar form, so that the solutions are more physically sound in the sense that the energy is no longer infinite and that the perturbation brought by the shock does not grow at large distances of it on the cold side in the soft potential case. 
\end{abstract}

\pacs{Valid PACS appear here}


\maketitle


\section{Introduction}

As well-known the inner structure of shock waves in simple gases is given by the solution of the Boltzmann kinetic equation with the appropriate boundary conditions far from the shock on the cold and hot side, supposing a plane shock. Therefore the study of the  solution of Boltzmann equations relevant for shocks makes a non trivial example of application of kinetic theory to realistic physical problems. In particular outside the neighborhood of Mach numbers slightly larger than $1$ perturbation methods are not directly applicable, as when computing the transport coefficients for instance.  Besides a difficult numerical approach of the solution of the Boltzmann equation to this problem, research has been directed over the years toward cases where results can be obtained in some limits. In this work we shall be concerned with the infinite Mach number limit. In this limit one looks at solution of 
spatially inhomogeneous and time independent Boltzmann equation. 
This very large Mach number limit was  studied in 1969 by Grad \cite{grad1969singular} for hard spheres. 
We shall write the Boltzmann equation as follows:  
\begin{equation}
\label{Boltzmann}
\xi_1 \partial_x f(x,\xi) \ = \ Q(f,f)(x,\xi),  \ \ \  x \in\mathbb{R}, \ \ \xi=(\xi_1,\xi_2,\xi_3) \in\mathbb{R}^3,
\end{equation}
in which, the collision operator takes the form
\begin{equation}
\label{BoltzmannCollision}
Q(f,f)(\xi) \ = \ \iint_{\mathbb{R}^3\times\mathbb{S}^2}B(\xi-\xi_*,\sigma)[f'_*f'-f_*f]\mathrm{d}\sigma \mathrm{d}\xi_*,
\end{equation}
where the notations $f$, $f_*$, $f_*'$ and $f'$ designate respectively the values $f(x,\xi)$, $f(x,\xi_*)$, $f(x,\xi_*')$ and $f(x,\xi')$ given
in terms of $\xi$, $\xi_*$ and $\sigma$ by the formulas
\begin{equation}
\label{Velocities}
\xi' \ = \ \frac{\xi+\xi_*}{2} + \frac{|\xi-\xi_*|}{2}\sigma, \ \ \ \ \ \xi'_* \ = \  \frac{\xi+\xi_*}{2} - \frac{|\xi-\xi_*|}{2}\sigma,
\end{equation}
with $\sigma\in\mathbb{S}^2$ being any arbitrary vector. The function $f(x, v)$ is the time independent distribution function, and depends on space (variable $x$) and of velocity (variable $v$). Being a probability distribution it must be positive and finite. 

The collision kernel $B(\xi-\xi_*,\sigma)$ depends on the solution of the two-body scattering problem and can be formulated as follows

\begin{equation}
\label{CollisionKernel}
\begin{aligned}
B(\xi-\xi_*,\sigma) \ & = \ B(\xi-\xi_*,\cos\theta)\ = \ b(\cos\theta)|\xi-\xi_*|^\gamma,\\
 \ \  \cos\theta \ & = \ \left\langle \frac{\xi-\xi_*}{|\xi-\xi_*|}\right\rangle,\end{aligned}
\end{equation}
where the exponent 
\begin{equation}
\label{gamma}
\gamma=\frac{s-5}{s-1}
\end{equation} is related to $s$,  $(s>2)$ which is minus the exponent for the assumed law of algebraic decay of the two-body forces , where $b$ is a locally smooth function given by the solution of the scattering problem with the 2-body potential and where the average is taken over all possible directions.

Grad assumed that, for a shock wave at infinite Mach number, the distribution of the gas has the following form: 
\begin{equation}
\label{GradDistribution}
f(x,\xi) \ = \ \alpha(x)\delta(\xi-c)  \ + \varphi(x,\xi),
\end{equation}
where $c=(u_0,0,0)$ (a constant vector) is the average speed of cold particles entering into the shock,  $\varphi(x,\xi)$ is the perturbation to the distribution function brought by the shock wave and $\delta(\xi-c)$ is a Dirac distribution of velocities keeping all velocities equal to the one on the cold side of the shock wave.  The physical idea behind
this decomposition of $f$ is that cold particles have a finite
probability to approach the shock at any large, finite
distance $x$ without collisions. These cold particles are
represented by distribution $\alpha(x)\delta(\xi-c)$, where $\alpha(x)$ is the number density at the distance $x$ from the shock. This function satisfies the boundary condition $\lim_{x\to -\infty}\alpha(x)=n$, at the cold side with $n$ number density of the cold gas. On the hot side $\lim_{x\to + \infty}\alpha(x)= 0$. The function $\varphi$ represents the particles having done one or more collisions and satisfies the boundary condition $\lim_{x\to -\infty}\varphi(x,\xi)=0$. The system for $\alpha$ and $\varphi$ reads very far from the shock on the cold side: 
\begin{equation}
\begin{aligned}
u_0\partial_x{\alpha} = &  -n\int_{\mathbb{R}^3}\varphi(\xi_*)|\xi_*-c|\mathrm{d}\xi_*,\\
\xi_1\partial_x\varphi  = &  n\iint_{\mathbb{R}^3\times\mathbb{S}^2}|\xi-\xi_*|[\varphi(\xi'_*)+\varphi(\xi')-\varphi(\xi)]\mathrm{d}\sigma\mathrm{d}\xi_*.
\end{aligned}
\end{equation}
Following Grad \cite{grad1969singular} for hard spheres $B(\xi-\xi_*,\sigma) $ has been taken as equal to $|\xi-\xi_*|$. The second equation is linearized because it applies to the cold side of the shock and far from it where the dominant contribution to the velocity distribution function is the delta function of the cold particles. This allows to linearize the equation for the small part of the distribution representing particles having made few collisions with the cold ones at large distances from the shock on this cold side. There the dominant contribution to the velocity distribution is the one of the cold particles having done no collision. 

Let us remark that the same strategy can also be applied to the Boltzmann-Nordheim equation, to obtain the system describing the interaction between the thermal clouds and the Bose-Einstein Condensates in finite temperature trapped Bose gases \cite{PomeauBinh}.

In \cite{pomeau1987shock}, the same problem for infinite total cross section of a power law interaction was studied by one of the authors. In this case, one cannot use the same method as Grad did for hard spheres since  the gain and loss
term of the Boltzmann collision operator would diverge separately proportional to an infinite cross section. However, it is natural to expect a continuous increase of the temperature on the cold side when approaching the
shock. As a result, an alternative approach was proposed where the delta distribution in \eqref{GradDistribution} is replaced by an approximated self-similar solution. The key idea is that there is an asymptotic solution for $x\to-\infty$ of \eqref{Boltzmann} having the following form
\begin{equation}
\label{PomeauSolution}
f(x,\xi) \ = \ |x|^\lambda G(v|x|^\lambda \mathrm{sgn}(x)),
\end{equation}
where $v=\xi-c$. Denote $v=(v_1,v_2,v_3)$, we then have $\xi_1=v_1+u_0$. However, the second order moment of this solution 
 cannot exist (see discussion around Eq.(\ref{PomeauEq}) below).  Being motivated by the interest in  solutions of the type \eqref{PomeauSolution}, Bobylev and Cercignani started a new
research direction on the study of the solutions of the space homogeneous Boltzmann equation, with a specific concern for those which do not have finite energy 
 \cite{bobylev2002self,bobylev2003eternal}. Note that infinite energy solutions do not have physical meaning in the present framework of kinetic theory of gases with collisions conserving the total kinetic energy, that is why it makes sense, as done in this paper, to look for a strategy of solution of the Boltzmann equation in shocks where the velocity distribution has finite energy.
 
 This made one problem for this approach by self-similar solutions of the decay of the perturbed velocity distribution on the cold side of the shock. As pointed out in \cite{pomeau1987shock}, there is also another problem with this idea of self-similar decay: if the exponent $s$ is smaller than $5$, the power of the ``decaying'' solution on the cold side becomes positive so that the solution so calculated does not decay but grows! Therefore there is need for improving this idea of self-similar solution decaying on the cold side of the shock. Specifically we look at the two problems. In section \ref{case 1}, we look at the case $s >5$ where the exponent of decay of the self-similar solution has the desired sign and yields a decaying solution far from the shock. This solution is consistent with the requirement of finite energy. We explain how to deal with this problem by adding another variable without breaking the self-similar structure of the solution but by getting rid of the energy problem. 
 
 Section  \ref{case 2} is devoted to the case $s< 5$ where the exponent of the self similar solution has the (wrong) positive sign. In this case we argue that the solution for the perturbation does not decay smoothly to infinity but has compact support in this direction and stops at finite distance from the shock.  

In section \ref{case 3} , we consider the case $s=5$ of Maxellian molecules. The corresponding  self-similar form is introduced and the relation with the a simplified spatial inhomogeneous Boltzmann equation is also discussed. 
 
 The general goal of this work is to provide a correction to the self-similar form \eqref{PomeauSolution}, so that the solutions are more physically sound in the sense that the energy is no longer infinite and that, if $s < 5$ the perturbation brought by the shock does not grow at large distances on the cold side.  We hope that those changes of the self-similar solutions would lead to some hints for the numerics of the shock waves coming from \eqref{Boltzmann}.

\section{Case 1: $s>5$}
\label{case 1} 
In the research for finite energy self-similar solutions of \eqref{Boltzmann} as $x\to- \infty$, let us consider the following  ansatz  
 \begin{equation}
\label{LogSolution1}
f(x,\xi) \ = \ |x|^{3\lambda} F(v|x|^\lambda \mathrm{sgn}(x),\lambda\beta \ln|x|),
\end{equation}
in which $\beta$ is any real constant. Since the function $F$  depends on the two quantities $v|x|^\lambda \mathrm{sgn}(x)$ and $\lambda\beta \ln|x|$, we then introduce the new variables $$v|x|^\lambda \mathrm{sgn}(x)=w$$ and $$\lambda\beta \ln|x|=\rho.$$

Inserting \eqref{LogSolution1} into \eqref{Boltzmann}, we obtain
 \begin{equation*}\begin{aligned}
& (u_0+v_1)[3\lambda |x|^{3\lambda-1} F \  + \ \lambda |x|^{4\lambda-1}\mathrm{sgn}(x)  v \cdot \partial_w F\\
& \ + \ \beta\lambda  |x|^{3\lambda-1} \partial_\rho F] =  \ Q[F,F]|x|^{\lambda\left(3 - \gamma\right)},\end{aligned}
\end{equation*}
where the differentiation in $w$ and the integration in the collision operator $Q[F,F]$ are with respect to the variable $w=v|x|^\lambda \mathrm{sgn}(x)$. Note that $\gamma$ is defined in \eqref{gamma}.

Rewriting the above equation in terms of $w=(w_1,w_2,w_3)$, we find
\begin{equation*}\begin{aligned}
(u_0+w_1|x|^{-\lambda}\mathrm{sgn}(x))[3\lambda |x|^{3\lambda-1} F \ & + \ \lambda |x|^{3\lambda-1}w \cdot \partial_w F \  \\
 \ \ \ \ \ \ \ + \ \beta\lambda  |x|^{3\lambda-1} \partial_\rho F] & \  =  \ Q[F,F]|x|^{\lambda\left(3 - \gamma\right)}.\end{aligned}
\end{equation*}
Note that for a fixed value of $w$, the term $u_0+w_1|x|^{-\lambda}\mathrm{sgn}(x)$ tends to $u_0$ as $x$ tends to $-\infty$ and $\lambda>0$. As a consequence, one could neglect the second term in the sum $u_0+w_1|x|^{-\lambda}\mathrm{sgn}(x)$ to have
\begin{equation*}
\begin{aligned}
u_0[3\lambda |x|^{3\lambda-1} F \ + \ \lambda |x|^{3\lambda-1}w \cdot \partial_w F \ & + \ \beta\lambda  |x|^{3\lambda-1} \partial_\rho F] \\
\indent & =   Q[F,F]|x|^{\lambda\left(3 - \gamma\right)}.\end{aligned}
\end{equation*}
Balancing powers of $|x|$, we arrive at
\begin{equation}
\label{LambdaValue1}
\lambda\gamma=1
\end{equation}
To make sense for the solution, this value of $\lambda$ must be positive: otherwise the self-similar solution increases far from the shock. The condition $\lambda >0$ imposes $s >5$. 

We finally obtain the equation
\begin{equation}\label{HardPotentailEq}
u_0[3\lambda  F \ + \ \lambda w \cdot \partial_w F \ + \ \beta\lambda  \partial_\rho F]\ =  \ Q[F,F].
\end{equation}
If we use the ansatz \eqref{PomeauSolution}, the following equation can be derived
\begin{equation}\label{PomeauEq}
u_0[3\lambda  F \ + \ \lambda w \cdot \partial_w F] =  \ Q[F,F].
\end{equation}
Multiplying both sides of \eqref{PomeauEq} with $w^2$, and integrating with respect to $w$, we find
\begin{equation}\label{Contradiction1}\int_{\mathbb{R}^3}3u_0\lambda F|w|^2 \mathrm{d}w \ + \ \int_{\mathbb{R}^3}u_0\lambda w \cdot \partial_w F w^2 \mathrm{d}w \ = \ 0,\end{equation}
which, after  integrating by parts the second term on the left hand side, implies 
\begin{equation}\label{Contradiction2}\int_{\mathbb{R}^3}3u_0\lambda F|w|^2 \mathrm{d}w \ - \ \sum_{i=1}^3\int_{\mathbb{R}^3}u_0\lambda  F \partial_{w_i} w_i|w|^2\mathrm{d}w \ = \ 0.\end{equation}
After rearranging the terms, the above equation can be written as
\begin{equation}\label{Contradiction3}-2\int_{\mathbb{R}^3}u_0\lambda F|w|^2 \mathrm{d}w \ = \ 0,\end{equation}
which leads to a contradiction. Notice that solutions of the form \eqref{PomeauSolution} motivated the study of infinite energy solutions \cite{bobylev2002self,bobylev2003eternal}.

However, thanks to the new variable $\rho$, we obtain 
$$\int_{\mathbb{R}^3}3u_0\lambda F|w|^2 \mathrm{d}w \ + \ \int_{\mathbb{R}^3}u_0\lambda w \cdot \partial_w F w^2 \mathrm{d}w$$
$$ \ + \  \lambda u_0 \beta \partial_\rho \int_{\mathbb{R}^3} F|w|^2 \mathrm{d}w  \ = \ 0,$$
leading to
\begin{equation}\label{Contradiction4} \lambda u_0 \beta \partial_\rho \int_{\mathbb{R}^3} F|w|^2 \mathrm{d}w \  -\ 2\int_{\mathbb{R}^3}u_0\lambda F|w|^2 \mathrm{d}w \ = \ 0,\end{equation}
which is no longer a contradiction and guarantees the boundedness of the energy of the solutions.

Notice that the above computation holds true for any choice of the constant $\beta$, which plays the role of a scaling parameter for the variable $\rho$. 

\section{Case 2: $s<5$}
\label{case 2} 

Besides the difficulty due to the conservation of energy, solved thanks to the introduction of the logarithmic variable $\rho$, the case $s< 5$ remains problematic because the self-similar stretching leads to a perturbation growing (instead of decaying, as it should) far from the shock on the cold side. This is to be changed to yield scaling laws in agreement with the expected behavior of the solution. A first indication in the direction of a possible solution comes from the remark that, if for $x$ large and negative, a solution decays like a negative power of $x$ and if this power becomes positive as a parameter changes, the positive exponent can be put in the expansion of a solution tending to zero at a finite value of $x$, called $x^*$ thereafter and taken negative. In other words the perturbation decaying like $(-x)^{- \lambda}$ with $\lambda$ positive at $x$ tends to minus infinity, becomes a function equal to zero for $(-x) > (-x^*)$ and behaving like $|x-x_*|^\lambda$ for $x < x^*$ and $|x-x_*|$ small. Because of the algebra giving the exponent $\lambda$ (see below) this exponent, as a function of $s$ keeps the same formal expression for $s$ bigger or smaller than $5$. In the limit case $s = 5$ the exponent is formally infinite and the solution decays exponentially (instead of algebraically) as $(-x)$ tends to infinity (see section \ref{case 3}). This makes a transition from an algebraic decay for $s > 5$ to a solution becoming exactly zero for $(-x) \geq (-x^*)$ 

Let us consider the following  ansatz  
 \begin{equation}
\label{LogSolution2}
f(x,\xi) \ = \ |x-x_*|^{3\lambda} F(v|x-x_*|^\lambda \mathrm{sgn}(x-x_*),\lambda\beta \ln|x-x_*|),
\end{equation}
in which $\beta$ is again  any real constant and $x_*$ is a fixed vector.  We again denote $$v|x-x_*|^\lambda \mathrm{sgn}(x-x_*)=w$$ and $$\lambda\beta \ln|x-x_*|=\rho.$$

Plugging \eqref{LogSolution2} into \eqref{Boltzmann}, we obtain also
 \begin{equation*}
 \begin{aligned}
&  (u_0+v_1)[3\lambda |x-x_*|^{3\lambda-1} F\\
&  +  \lambda |x-x_*|^{4\lambda-1}\mathrm{sgn}(x-x_*) v \cdot \partial_w F \\
&\indent\indent \ + \ \beta\lambda  |x-x_*|^{3\lambda-1} \partial_\rho F] \ =  \ Q[F,F]|x-x_*|^{\lambda\left(3 - \gamma\right)},
\end{aligned}
\end{equation*}
where $\gamma$ is defined in \eqref{gamma}.
In terms of $w$, the equation can be rewritten as
 \begin{equation*}
 \begin{aligned}
& (u_0+w_1|x-x_*|^{-\lambda}\mathrm{sgn}(x-x_*))[3\lambda |x-x_*|^{3\lambda-1} F \\
&\indent +  \lambda |x-x_*|^{3\lambda-1}w \cdot \partial_w F \\
 &\indent +  \beta\lambda  |x-x_*|^{3\lambda-1} \partial_\rho F]  =   Q[F,F]|x-x_*|^{\lambda\left(3 - \gamma\right)}.
\end{aligned}
\end{equation*}
Now, the term $u_0+w_1|x-x_*|^{-\lambda}\mathrm{sgn}(x-x_*)$ tends to $u_0$ as $x$ tends to $x_*$ and $\lambda<0$. As a consequence, we have
\begin{equation*}\begin{aligned}
& u_0[3\lambda |x-x_*|^{3\lambda-1} F   +   \lambda |x-x_*|^{3\lambda-1}w \cdot \partial_w F    \\
& +  \beta\lambda  |x-x_*|^{3\lambda-1} \partial_\rho F]  =   Q[F,F]|x-x_*|^{\lambda\left(3 - \gamma\right)}.\end{aligned}
\end{equation*}
Balancing power of $|x-x_*|$, we obtain
\begin{equation}
\label{LambdaValue1}
\lambda=\frac{1}{\gamma}<0.
\end{equation}
We finally have the equation
\begin{equation}\label{SoftPotentailEq}
u_0[3\lambda  F \ + \ \lambda w \cdot \partial_w F \ + \ \lambda \beta \partial_\rho F] \ =  \ Q[F,F].
\end{equation}
Again, thanks to the new variable $\rho$, the boundedness of the energy density of the solutions is guaranteed.
\section{Case 3: $s = 5$}
\label{case 3} 

For Maxwell molecules, we could use the ansatz
 \begin{equation}
\label{MaxwellSol}
f(x,\xi) \ = \ e^{3\lambda x}F(v e^{\lambda x},\beta \lambda x),
\end{equation}
and define $$w=v e^{\lambda x},\ \ \ \rho=\beta \lambda  x.$$
The same argument as above also leads to the same equation 
\begin{equation}\label{SoftPotentailEqa}
\lambda (u_0+v_1)[3  F \ + \  w \cdot \partial_w F \ + \ \beta   \partial_\rho F] \ =  \ Q[F,F],
\end{equation}
which then leads to 
\begin{equation}\label{SoftPotentailEqb}
\lambda (u_0+w_1 e^{-\lambda x})[3  F \ + \  w \cdot \partial_w F \ + \ \beta   \partial_\rho F] \ =  \ Q[F,F].
\end{equation}
Dropping $w_1 e^{-\lambda x}$ in the factor $u_0+w_1 e^{-\lambda x}$, we finally arrive at
\begin{equation}\label{SoftPotentailEq}
\lambda u_0[3  F \ + \  w \cdot \partial_w F \ + \ \beta   \partial_\rho F] \ =  \ Q[F,F].
\end{equation}
Notice that in this case $\lambda$ does not have an explicit value, as given in Eq.\eqref{LambdaValue1}. On the other hand, equation \eqref{SoftPotentailEq} looks similar to an eigenvalue problem. However, different from classical eigenvalue problems, the right hand side of \eqref{SoftPotentailEq} is nonlinear and the left hand side of \eqref{SoftPotentailEq} involves a transport process. This {\it  nonlinear eigenvalue problem} could result in multiple values of $\lambda$ as  nonlinear eigenvalues or there could be no  nonlinear eigenvalue at all. 

In the next step, we will start from a completely different  equation, then introduce a self-similar form for the solution of this equation, and derive \eqref{SoftPotentailEq} in this new context. Let us consider a function $K(t,\tau,\vartheta)$, being the solution of the Boltzmann equation
\begin{equation}
\label{TwoTimeBoltzmann}
\partial_t K \ + \ \partial_\tau K \ = \ Q[K,K]. 
\end{equation}
In the above equation $t\in[0,\infty)$ is the time variable, $\tau\in(-\infty,\infty)$ is the space variable and $\vartheta\in\mathbb{R}^3$ is the velocity variable.
Suppose that $K$ takes the self-similar form 
\begin{equation}
\label{NewSelfsimilarity}
K(t,\tau,\vartheta)\ = \ e^{3u_0\lambda t}H(\beta\lambda u_0\tau, \vartheta  e^{u_0\lambda t}),
\end{equation}
and denote 
\begin{equation}
\label{newvariable}
\bar{w}\ =\ \vartheta  e^{u_0\lambda t}, \ \ \ \ \bar\rho=\beta\lambda u_0\tau.
\end{equation}
Plugging this anszart into \eqref{TwoTimeBoltzmann}, we arrive at
\begin{equation}
\label{TwoTimeBoltzmann1}\begin{aligned}
e^{3u_0\lambda t}3u_0\lambda  H \ + \ e^{4u_0\lambda t}u_0\lambda   \vartheta  \cdot\partial_{\bar{w}}H \ & + \ \beta u_0\lambda e^{3u_0\lambda t}\partial_{\bar\rho} H\\
 \ &  = \ e^{3u_0\lambda t}Q[H,H],\end{aligned}
\end{equation}
leading to
\begin{equation}
\label{TwoTimeBoltzmann1}
\lambda u_0[3  H \ + \  \bar{w}\cdot \partial_{\bar{w}} H \ + \ \beta   \partial_{\bar\rho} H] \ =  \ Q[H,H],
\end{equation}
which has exactly the same formulation as Equation \eqref{SoftPotentailEq}, where $\bar{w}$ and $\bar\rho$ play the roles of $w$ and $\rho$. 
As a result, starting from an unrelated equation \eqref{TwoTimeBoltzmann}, by a self-similar argument, we can still obtain \eqref{TwoTimeBoltzmann1}. Thus,  to study \eqref{SoftPotentailEq}, one possibility is to  study the {\it two-time spatial homogeneous Boltzmann equation for Maxwell molecules} \eqref{TwoTimeBoltzmann} instead.  In \eqref{TwoTimeBoltzmann}, the first time variable $t$ belongs to $\mathbb{R}_+$ and the second time variable $\tau$ belongs to $\mathbb{R}$.

The operator $\partial_t K \ + \ \partial_\tau K$ is a transport operator, with $\tau$ being the one-dimensional spatial variable. As a result, equation \eqref{TwoTimeBoltzmann} is a {\it simplified spatial inhomogeneous Boltzmann equation}, in which the coefficient associated to the term $\partial_\tau K$ in the transport operator is one. 

In any case,  to be physically sound, one could look for solutions of \eqref{TwoTimeBoltzmann}  with finite energy. 
\section{Summary and conclusion}
\label{Summaryandconclusion} Besides the (unrealistic) case of hard spheres, the way the velocity distribution behaves on the cold side of shocks at infinite Mach number  was unknown for realistic potential with an infinite total cross section. This was for two reasons. For hard potentials ($s >5$) the equation for the self-similar decay was unable to satisfy the conservation of energy. For soft potentials  ($s < 5$) the self-similar solution does not even decay far from the shock. This paper introduces three new ansatz permitting to circumvent both difficulties. First the energy problem is eliminated thanks to the introduction of another logarithmic variable in the similarity  assumption. Then, in the case of soft potentials, it is shown that the perturbation brought by the shock on the cold side stops exactly at a finite distance from the shock. Indeed our derivation is not supported by detailed mathematical stimulates, but it yields at least a coherent schema for the solution of an interesting problem in kinetic theory.

{\bf Acknowledgements.} 
 M.-B. Tran is partially supported by NSF Grant DMS-1814149 and NSF Grant DMS-1854453.


\def\cprime{$'$}

\end{document}